\documentclass[12pt]{article}
\begin{document}
\pagestyle{plain}
\setcounter{page}{1}
\begin{center}
{\large \textbf{Spontaneous Violation of Lorentz Invariance and
Ultra-High Energy Cosmic Rays}}
\vskip 0.2 true in
{\large J. W. Moffat}\footnote{e-mail: john.moffat@utoronto.ca}
\vskip 0.2 true in
\textit{Perimeter Institute for
Theoretical Physics, Waterloo, Ontario N2J 2W9, Canada}
\vskip 0.1 true in
and
\vskip 0.1 true in
\textit{Department of Physics, University of
Toronto, Toronto, Ontario M5S 1A7, Canada}

\vskip 0.2 true in
\begin{abstract}
We propose that local Lorentz invariance is spontaneously
violated at high energies, due to a nonvanishing vacuum expectation value of a
vector field $\phi^\mu$, as a possible explanation of the observation of ultra-high
energy cosmic rays with an energy above the GZK cutoff. Certain consequences of
spontaneous breaking of Lorentz invariance in cosmology are discussed.
\end{abstract}
\vskip 0.3 true in \end{center}

\date{\today}

\section{\bf Introduction}

The possibility that cosmic ray protons may interact with the
photons of the cosmic microwave background (CMB) with an energy
larger than the Greisen-Zatsepin-Kuzmin (GZK) cutoff, $5\times 10^{10}$ GeV
~\cite{Zatsepin}, such that the center-of-mass energy can be larger than the
threshold for pion photoproduction, has been the topic of discussion
recently~\cite{Camelia,Piran,Glashow,Camelia2,Smolin,Toller}.
Experiments have detected cosmic rays with significantly larger
energies~\cite{Takeda,Stecker}, although arguments have recently been made that the
GZK cutoff may already have been discovered~\cite{Bahcall}. It has been proposed
that the dispersion relation which connects the energy $p^0=E/c$ and the momentum
${\vec p}$ be modified
\begin{equation}
\label{dispersion}
F(E)\not= c[({\vec p})^2+m^2c^2]^{1/2}.
\end{equation}
The function $F$ contains a new high-energy scale which could be
associated with properties of quantum gravity~\cite{Camelia,Smolin,Kifune,Majid}.

The modified dispersion relation (\ref{dispersion}) is
rotationally invariant but breaks Lorentz invariance. Various proposals have been
put forward to describe Lorentz violation. A phenomenological framework for
analyzing possible departures from Lorentz invariance was given by Coleman and
Glashow~\cite{Glashow2}, who assumed that Lorentz invariance is broken
perturbatively in standard model quantum field theory. It was
assumed further that the action is rotationally invariant in a
preferred frame, which is the rest frame of the cosmic microwave
background (CMB). In this preferred frame (\ref{dispersion}) is
valid, while in other frames it is not rotationally invariant.
Another possibility is to retain special relativity in a
modified form i.e. the dispersion relation is valid in all
inertial frames, but the linear Lorentz transformation laws are
modified~\cite{Camelia2,Smolin}. This proposal is sometimes
called ``double special relativity'', for the symmetry
transformations possess two universal constants, the speed of
light and a fundamental energy scale which is usually identified
with the Planck energy $E_P\sim 10^{19}$ GeV. It has been argued
that the latter proposal implies that the conservation of
energy-momentum is not conserved at high-energies~\cite{Toller}.

An alternative solution to the initial value
problems of cosmology, based on a variable speed of light (VSL)
~\cite{Moffat,Moffat2}, postulated that in the
very early universe at a time $t\sim t_P\sim 10^{-43}$ sec.,
where $t_P$ denotes the Planck time, the local Lorentz
invariance of the ground state of the universe was spontaneously
broken by means of a non-zero vacuum expectation value (vev) of a
vector field, $\langle\phi^\mu\rangle_0 \not=0$. At a temperature
$T < T_c$, the local Lorentz
symmetry of the vacuum was restored corresponding to a
``non-restoration'' of the symmetry group $SO(3,1)$.  Above
$T_c$ the symmetry of the ground state of the universe was
broken from $SO(3,1)$ down to $SO(3)$, and the domain formed by
the direction of $\langle\phi^\mu\rangle_0$ produced an arrow of
time pointing in the direction of increasing entropy and the
expansion of the universe \footnote{For alternative VSL models of cosmology, see
references ~\cite{Clayton,Clayton2,Magueijo}.}.

The notion that as the temperature of the universe increases, a larger
symmetry group $SO(3,1)$ can spontaneously break to a smaller group $SO(3)$
seems counter-intuitive. Heating a superconductor restores gauge
invariance, and heating a ferromagnet restores rotational invariance.
Non-restoration would appear to violate the second law of
thermodynamics. This, however, is not the case, for certain
ferroelectric crystals such as Rochelle or Seignette salt,
possess a smaller invariance group above a critical temperature,
$T=T_c$, than below it~\cite{Jona} \footnote{Rochelle salt
possesses a lower Curie point at $-18^o\,C$, below which the
Rochelle crystal is orthorhombic and above which it is
monoclinic.}. Explicit models of 4-D field theories have been
constructed in which the non-restoration of symmetries
occurs at high temperatures~\cite{Weinberg}.

In the following, we shall postulate that local Lorentz invariance of the vacuum
state is spontaneously broken, so that $SO(3,1)\rightarrow SO(3)$ at some energy
$E_c$. The model is based on a Maxwell-Proca action for a vector field $\phi^\mu$,
which avoids instabilities and the existence of negative energy ghost
states~\cite{Isenberg,Clayton}. We attempt to emulate the successes of the standard
model of particle physics~\cite{Cheng}, in which ``soft'', spontaneous breaking of
the internal symmetries by a Higgs mechanism plays a crucial role. In our scenario,
local Lorentz symmetry at low energies is simply an accident of nature, i.e. the
ground state happens to be found at low energies in a particular local Lorentz
invariant vacuum state, and transitions away from this vacuum state may well happen
at high energies \footnote{Gravity theories with a preferred frame have been
considered by Gasperini~\cite{Gasperini} and Jacobson and Mattingly~\cite{Jacobson}.
In these models, the preferred frame was described by a unit timelike vector
(``aether'') and was a {\it permanent} feature of spacetime, and not related to a
broken and subsequently restored symmetry phase of spacetime. A model of spontaneous
breaking of Lorentz invariance inspired by string theory was constructed by
Kostelecky and Samuel~\cite{Samuel,Kostelecky,Lehnert}, and used to obtain
experimental bounds on possible Lorentz violations in the standard model and in
electrodynamics.}.

We shall base our formalism on four-dimensional Einstein gravity,
anticipating a possible basis for a future quantum gravity theory in which the
spontaneous violation of local Lorentz invariance plays a fundamental role
\footnote{In previous work~\cite{Moffat,Moffat2}, we used a vierbein and
spin-gauge connection to describe the spontaneous breaking of local Lorentz
invariance. In the following, we shall use the conventional metric
$g_{\mu\nu}$ and Christoffel connection $\Gamma^\lambda_{\mu\nu}$ to depict the
pseudo-Riemannian geometry.}. In certain
spin-foam and loop models of quantum gravity~\cite{Smolin,Oriti}, discrete spacetime
structure at the Planck length, $L_P\sim 10^{-33}$ cm, leads to a violation of
Lorentz invariance. When a cutoff is introduced into a quantum field theory or in
perturbative gravitational calculations of loop graphs, Lorentz invariance is
violated. Therefore, it has long been suspected that if a consistent way of breaking
Lorentz invariance could be discovered, it would lead to a finite quantum field
theory and quantum gravity theory.

Investigations of possible trans-Planckian energy modifications of the calculations
of the primordial fluctuation spectrum have used the modified dispersion relation
(\ref{dispersion}), due to the possibility that the initial tiny comoving wave modes
born in the quantum ground state of the inflaton field could occur at energies above
the Planck energy $E_P$~\cite{Brandenberger}.

The total action is given by
\begin{equation}
S=S_G+S_M+S_\phi,
\end{equation}
where
\begin{equation}
\label{grav}
S_G=\frac{1}{\kappa}\int dtd^3x\sqrt{-g}(R+2\Lambda),
\end{equation}
and $g={\rm det}(g_{\mu\nu})$, $\kappa=16\pi G/c^4$, $\Lambda$ is the
cosmological constant and $S_M$ is the matter action. Moreover,
\begin{equation}
\label{Proca}
S_\phi=\int
dtd^3x\sqrt{-g}\biggl[-\frac{1}{4}g^{\mu\nu}g^{\alpha\beta}B_{\mu\alpha}B_{\nu\beta}
+V(\phi)\biggr],
\end{equation}
where $B_{\mu\nu}=\partial_\mu\phi_\nu-\partial_\nu\phi_\mu$. The potential $V(\phi)$ is
\begin{equation}
V(\phi)=\frac{1}{2}\mu^2\phi^\mu\phi_\mu+W(\phi),
\end{equation}
where $W(\phi)$ is a function of $\phi^\mu\phi_\mu$. The total action $S$ is
diffeomorphism and locally Lorentz invariant.

The energy-momentum tensor for matter is
\begin{equation}
T_{M\mu\nu}=-\frac{2}{\sqrt{-g}}\frac{\delta S_M}{\delta
g^{\mu\nu}},
\end{equation}
and we have
\begin{equation}
T_{\phi\mu\nu}=B_{\mu\rho}B^\rho_\nu-\frac{1}{4}g_{\mu\nu}g^{\rho\sigma}
g^{\lambda\tau}B_{\rho\lambda}B_{\sigma\tau}-2\frac{\partial V(\phi)}
{\partial g^{\mu\nu}}+g_{\mu\nu}V(\phi).
\end{equation}

We observe that although the Maxwell-Proca action (\ref{Proca})
is not invariant under Abelian $U(1)$ gauge transformations,
{\it the quantized version of the action is fully consistent and
free of negative energy instabilities}~\cite{Isenberg}. A Hamiltonian
constraint analysis reveals that
the time component $\phi^0$ satisfies a constraint equation
and does not physically propagate, thereby, avoiding negative energy states. This
solves the problem of obtaining a physically consistent model of symmetry breaking
for the local Lorentz group $SO(3,1)$, which is a non-compact group \footnote{For
example, if we had chosen the kinetic part of the action for $\phi^\mu$ to be
\begin{equation}
S_{\phi{\rm kin}}=\frac{1}{2}\int
dtd^3x\sqrt{-g}\partial_\mu\phi^\alpha\partial^\mu\phi_\alpha,
\end{equation}
then for the compact group $SO(4)$ there will not be any negative
energy (ghost) modes, for $SO(4)$ is a compact group, whereas
for the noncompact group $SO(3,1)$ negative energy modes will cause the kinetic
energy contribution to be unstable and the Hamiltonian will not be bounded from
below.}.

The field equations obtained from the action principle are given
by
\begin{equation}
G_{\mu\nu}=\frac{\kappa}{2}T_{\mu\nu}+g_{\mu\nu}\Lambda,
\end{equation}
\begin{equation}
\label{Bequation}
-\nabla_\nu B^{\mu\nu}+\frac{\partial V(\phi)}{\partial\phi^\mu}=0.
\end{equation}
We have $G_{\mu\nu}=R_{\mu\nu}-\frac{1}{2}g_{\mu\nu}R$
and
\begin{equation}
T_{\mu\nu}=T_{M\mu\nu}+T_{\phi\mu\nu}.
\end{equation}
The Bianchi identities $\nabla_\nu G^{\mu\nu}=0$ lead to the
conservation law: $\nabla_\nu T^{\mu\nu}=0$.

Let us now choose $V(\phi)$ to be of the form of a ``Mexican hat''
potential
\begin{equation}
\label{phipotential}
V(\phi)=-\frac{1}{2}\mu^2\phi_\mu\phi^\mu+\frac{1}{4}\lambda(\phi_\mu\phi^\mu)^2+V_0,
\end{equation} where the coupling constant $\lambda > 0$, $\mu^2 >0$, and
the potential is bounded from below. If $V$ has a minimum at
\begin{equation}
\phi_\mu\equiv v_\mu=\langle\phi_\mu\rangle_0,
\end{equation}
then the spontaneously broken solution is
given by
\begin{equation}
\phi^\mu\phi_\mu\equiv v^2=\mu^2/\lambda,
\end{equation}
where $v^2=(v_\mu)^2$.  This manifold of points of the minima of $V(\phi)$ is
invariant under local Lorentz transformations
\begin{equation}
\phi^{'\mu}=\Lambda^\mu_\nu\phi^\nu.
\end{equation}

We can choose the ground state to be described by the timelike
vector
\begin{equation}
\label{vev}
\phi_\mu^{(0)}=\delta_{\mu0}v=\delta_{\mu0}(\mu^2/\lambda)^{1/2}.
\end{equation}
The homogeneous Lorentz group $SO(3,1)$ is broken down to the
spatial rotation group $SO(3)$. The three rotation generators $J_i\,(i=1,2,3)$ leave
the vacuum invariant, $J_iv_i=0$, while the three Lorentz-boost generators $K_i$
break the vacuum symmetry, $K_iv_i\not= 0$.

We shall consider the perturbations about the stable vacuum state, $\chi(x),
\theta^i(x)$ (i=1,2,3):
\begin{equation}
\phi^0=v+\chi, \quad \phi^i=\theta^i.
\end{equation}
The potential term in the action for the perturbations in a local, flat patch of
spacetime, for which $g_{\mu\nu}\approx\eta_{\mu\nu}$ (where $\eta_{\mu\nu}={\rm
diag}(1,-1,-1,-1)$ is the Minkowski spacetime metric) has the form
\begin{equation}
V(\phi)=-\frac{\mu^2}{2}[(v+\chi)^2-(\theta^i)^2]
+\frac{\lambda}{4}\lambda[(v+\chi)^2-(\theta^i)^2]^2+V_0,
\end{equation} and the action becomes
\begin{equation}
S_\phi=\int dtd^3x\biggl\{\frac{1}{4}[-(\partial_i\theta_j-\partial_j\theta_i)^2
+2(\partial_0\theta_i-\partial_i\chi)^2]+\frac{\mu^2}{2}\chi^2
$$ $$
+{\rm cubic\, and\, quartic\,
terms\, in\, \chi,\theta}+{\rm const.}\biggr\}.
\end{equation}
We see that the action of the perturbations is invariant under $SO(3)$ rotations
of the fields $\theta^i$, since it only contains the combination $(\theta^i)^2$,
but it does not possess the full Lorentz $SO(3,1)$ symmetry invariance. It follows
that $\theta^i$ are three massless Nambu-Goldstone fields.

The vector field $\phi^\mu$ perturbations have three massless Nambu-Goldstone modes
that could produce long-range fifth force effects, when $\phi^\mu$ interacts with
matter. Fifth force experiments have not detected any observable effects, so this
will put a bound on the strength of the coupling of $\phi^\mu$ with
matter~\cite{Fischbach}.

\section{\bf Spontaneously Broken Lorentz Invariance and Cosmology}

In the spontaneously broken phase of the evolution of the universe, the
spacetime manifold has been broken down to $R\times SO(3)$. The
three-dimensional space with $SO(3)$ symmetry is assumed to be the
homogeneous and isotropic FRW solution:
\begin{equation}
d\sigma^2=R^2(t)\biggl[\frac{dr^2}{1-kr^2}+r^2(d\theta^2+\sin^2\theta
d\phi^2)\biggr],
\end{equation}
where $k=0,+1,-1$ corresponding to a flat, closed and open universe,
respectively, and $t$ is the external time variable. This describes the
space of our ground state in the symmetry broken phase and it has
the correct subspace structure for our FRW universe with the metric
\begin{equation}
\label{metric}
ds^2\equiv
g_{\mu\nu}dx^\mu
dx^\nu=dt^2c^2-R^2(t)\biggl[\frac{dr^2}{1-kr^2}+r^2(d\theta^2+\sin^2\theta
d\phi^2)\biggr].
\end{equation}
The Newtonian ``time'' $t$ is the {\it cosmological time} measured by
standard clocks.

The Friedmann equations are
\begin{equation}
H^2+\frac{kc^2}{R^2}=\frac{8\pi G\rho}{3}+\frac{\Lambda c^2}{3},
\end{equation}
\begin{equation}
{\ddot R}=-\frac{4\pi G}{3}(\rho+3\frac{p}{c^2})R+\frac{\Lambda c^2}{3}R,
\end{equation}
where $H={\dot R}/R$ and $\rho=\rho_M+\rho_\phi$ and $p=p_M+p_\phi$.

In our homogeneous and isotropic cosmology, we have
$B_{\mu\nu}=0$ and $\phi_i=0$ (i=1,2,3).
We have from Eq.({\ref{Bequation}) that
$dV(\phi)/d\phi^0=0$ and from (\ref{phipotential}), we find that at the
minimum of the potential: $\phi_0\equiv
\langle\phi\rangle_0=(\mu^2/\lambda)^{1/2}=v$, which is the result
(\ref{vev}) obtained previously for our choice of gauge and
spontaneous symmetry breaking. The energy momentum tensor for the
$\phi_\mu$ field is now given by
\begin{equation}
\label{cosenergy}
T_{\phi\mu}^\nu=\delta^\nu_\mu V(\phi),
\end{equation}
where
\begin{equation}
V(\phi)=-\frac{\mu^4}{4\lambda}+V_0.
\end{equation}

For the fluid in comoving coordinates, we have
\begin{equation}
T_{\phi\mu}^\nu={\rm diag}(\rho_\phi c^2, -p_\phi, -p_\phi, -p_\phi).
\end{equation}
We obtain from (\ref{cosenergy}):
\begin{equation}
T^0_{\phi0}\equiv\rho_\phi
c^2=V(\phi),\quad  T_{\phi i}^j\equiv -p_\phi\delta_i^j=\delta_i^jV(\phi),
\end{equation}
which gives the equation of state, $p_\phi=-\rho_\phi c^2$, associated with the
effective cosmological constant
\begin{equation}
\Lambda_{\rm eff}=\Lambda+\Lambda_\phi.
\end{equation}

If $\Lambda=0$ and the vacuum density $\rho_\phi$ dominates over the matter
density $\rho_M$, then in the spontaneously broken phase the universe will
be inflationary with ${\ddot R}>0$ and be dominated by a de Sitter
solution~\cite{Linde}. If the period of spontaneous symmetry breaking
lasts long enough to generate $\sim 60$-efolds of inflation, then it
solves the horizon and flatness problems and can generate a scale
invariant fluctuation spectrum. The inflationary period ends when the
local $SO(3,1)$ symmetry is restored.

\section{\bf The GZK Cutoff and Conclusions}

We have shown that by beginning with a diffeomorphism invariant action that includes
the Einstein-Hilbert action, due to a nonvanishing vev
$\langle\phi^\mu\rangle_0\not=0$, the local Lorentz invariance of the vacuum state is
spontaneously broken at high energies. In a small patch of flat spacetime the
homogeneous Lorentz group $SO(3,1)$ is broken down to the homogeneous rotation group
$SO(3)$. In the spontaneously broken phase at high energies, the modified dispersion
relation:
\begin{equation}
E^2-c^2({\vec p})^2-m^2c^4=f(E^2),
\end{equation}
is invariant under the subgroup of rotations $SO(3)$, but is not locally
Lorentz invariant. In this scenario, a possible modified dispersion relation would
correspond to choosing
\begin{equation}
f(E^2)=\eta E^2\biggl(\frac{E}{E_P}\biggr)^\alpha\sim \eta c^2({\vec
p})^2\biggl(\frac{E}{E_P}\biggr)^\alpha,
\end{equation}
where $\eta$ is a dimensionless parameter of order 1. This modified dispersion
relation~\cite{Camelia,Piran} has been
related to theories of quantum gravity~\cite{Camelia,Smolin,Kifune}. For
$\alpha\sim 1-2$, this deformed dispersion relation will show a significant upward
shift above the GZK cutoff determined by
\begin{equation}
E\epsilon=\frac{c^4[(m_{\rm prot}+m_\pi)^2-m^2_{\rm prot}]}{4},
\end{equation}
where $E$ and $\epsilon$ denote the energy of the proton and the photon,
respectively, in the photoproduction process $p+\gamma\rightarrow p+\pi$.

In the Lorentz invariant deformed models of dispersion relations, or the ``doubly
special relativity'' models~\cite{Camelia2}, it is not clear how one incorporates
gravity theory, and, in particular, Einstein's theory of gravity.  In contrast, in
our model the initial action contains the Einstein-Hilbert gravity theory. An
important feature of the Maxwell-Proca action $S_\phi$ is that the vector field
$\phi^\mu$ can be consistently quantized, even though the action is not gauge
invariant. It is hoped that future experiments will confirm whether there is an
ultra-high energy cosmic ray ``threshold anomaly'' associated with the GZK cutoff.

The spontaneously broken symmetry model described here could be important for
early universe inflationary and VSL cosmology~\cite{Linde,Moffat,Moffat2}.

\vskip0.2 true in
\textbf{Acknowledgments}
\vskip0.2 true in
This work was supported by the Natural Sciences and Engineering Research Council of
Canada. I thank Michael Clayton and Pierre Savaria for helpful discussions.
\vskip0.5 true in

\end{document}